\documentclass{cernrep}
\usepackage[title]{appendix}
\usepackage{amsmath,amssymb,bm}
\usepackage{authblk}
\newcommand{\braket}[1]{\langle {#1} \rangle }
\newcommand{\ket}[1]{|{#1} \rangle }

\usepackage{amsmath}	
\usepackage{color}

\begin{document}
	\title{One-- and two-- neutron halo at the dripline. From $^{11}$Be to $^{11}$Li and back: $^{10}$Li and parity inversion.}
	\author[1,2]{R. A. Broglia}
	\author[3]{F. Barranco}
	\author[4]{G. Potel}
	\author[5]{E. Vigezzi}
	\affil[1]{Dipartimento di Fisica, Universit\`a degli Studi di Milano,
		Via Celoria 16, 
		I-20133 Milano, Italy}
	\affil[2]{The Niels Bohr Institute, University of Copenhagen, 
		DK-2100 Copenhagen, Denmark}
	\affil[3]{Departamento de F\`isica Aplicada III,
		Escuela Superior de Ingenieros, Universidad de Sevilla, Camino de los Descubrimientos, 	Sevilla, Spain}
	\affil[4]{National Superconducting Cyclotron Laboratory, Michigan State University, East Lansing, Michigan 48824, USA}
	\affil[5]{INFN Sezione di Milano, Italy}
	\date{\today}
	\maketitle
	\begin{abstract}
	The nuclei $^{11}$Be and $^{11}$Li provide paradigmatic examples of one--and two--neutron
	halo systems. Because the reaction $^1$H($^{11}$Li,$^9$Li)$^3$H is dominated by successive transfer, one can use the quantitative picture emerging from a nuclear field theory description of the structure and reaction mechanism of the above Cooper pair transfer
	process and of the $^2$H($^{10}$Be,$^{11}$Be)$^1$H and $^1$H($^{11}$Be, $^{10}$Be)$^2$H reactions, to shed light on
	the structure of $^{10}$Li. This analysis provides important support for a parity inverted	scenario with a $1/2^+$ virtual state at about 0.2 MeV.
		\end{abstract}
\section{Introduction}
Potential energy thrives on relative fixed positions of particles, fluctuations on delocalization.  A quantitative measure of these two contrasting effects is provided in many--body systems, by the quantality parameter $q=\frac{\hbar^2}{ma^2}\frac{1}{|v_0|}$, where $a$	and $v_0$ are the interaction range and strength, respectively. In the nuclear case $q\approx0.5$ ($a=0.9$ fm, $v_0=-100$ MeV), testifying to a quantal--fluctuation--dominated regime and thus delocalization\footnote{It is of notice that in the above reasoning no reference to the Pauli principle was made. The fact that $q\ll1$ implies fixed positions while $q\approx 1$ delocalization, is essentially independent of the statistics obeyed by the particles.} which can be described at profit in terms of a mean field, shells, and magic numbers.
\section{Neutron drip lines}
If neutrons are progressively added to a light normal nucleus, Pauli principle forces the system,  when the core becomes neutron saturated, to expel most of the wavefunction of the last neutrons outside to form a halo which, because of its large size, can have lower momentum. It is an open question how nature stabilizes such fragile objects and provides the glue to bind the halo neutrons to the core.
Within this context, the fact that   $^9_3$Li$_6$, $^{10}_3$Li$_7$, $^{11}_3$Li$_8$  are bound (closed shell (cs) in neutrons), unbound (one--neutron outside cs), barely bound (Cooper pair outside cs) nuclei respectively, provides evidence of a pairing mechanism resulting in a Cooper pair halo.
\subsection{One-neutron halo}
To elaborate on this issue, use is made of the bound ($T_{1/2}=13.76$ s) one--neutron halo nucleus outside the $N=6$ closed shell, $^{11}$Be. To create a halo system one has to have an $s_{1/2}$--level (no centrifugal barrier) at threshold. But then, why not retain $N=8$ as a magic number and eventually $^{13}$Be as one--neutron halo? Because one has to bring the $s_{1/2}$ down in energy to become weakly bound. And to do so one has essentially one possibility. To dress the bare $2s_{1/2}$ state ($\epsilon_{s_{1/2}}\approx0.07$ MeV) with the quadrupole vibration of the core, a process which binds the dressed level by about 0.5 MeV. But equally inescapable is the weakening of the binding energy of the $p_{1/2}$ orbital ($\epsilon_{p_{1/2}}\approx-3.04$ MeV) to an essentially threshold situation ($\tilde\epsilon_{p_{1/2}}\approx-0.18$ MeV). This is a result of the fact that the amplitude of the main neutron component of the quadrupole vibration namely the $(p_{1/2},p^{-1}_{3/2})_{2^+}$ is close to one. Pauli principle between the odd $p_{1/2}$ neutron and the same particle involved in the collective mode leads to almost 3 MeV repulsion (+2.86 MeV). In other words, to make the $2s$--state barely bound so as to produce a one--neutron halo, one is forced at the same time to weaken conspicuously the binding of the $p_{1/2}$ state \cite{Barranco:17}. While parity inversion is not a condition, what is inescapable is the melting away of the $N=8$ magic number and the appearance of the $N=6$ closed shell. Also of an $E1$--transition between the parity inverted levels carrying essentially one $B_W$ (Weisskopf unit) (see App. \ref{appB}). A consequence of the very poor overlap existing between halo neutron and core nucleons which impedes the GDR to depopulate the low--energy $1/2^+\to1/2^-$ transition, being forced to leave about 10\% of the TRK sum rule anchored to this low--lying $E1$--transition.
\subsection{Two--neutron halo}\label{S2.2}
While it will be natural, within the above scenario, to deal with the unbound system $^{10}$Li, it is likely more useful to start with the bound ($T_{1/2}=8.5$ ms) two--neutron halo nucleus $^{11}$Li \cite{Barranco:01}. Simple estimates of such neutron halo Cooper pair can be made by assuming that the calculation scheme used in $^{11}$Be \cite{Barranco:17}, and based on  the values of $\beta_2$, $\hbar\omega_2$ (core), $|E_{corr}|\approx S_{1n}$ and 10\% TRK, is  transferable to $^{11}$Li and eventually through it, to the virtual system $^{10}$Li (see Section \ref{A3}, App. \ref{AppA} below). The physics at the basis of this ansatz is quite general and operative: a) poor overlap between halo neutrons and core nucleons (thus low--energy presence of a substantial fraction of the TRK sum rule); b) Lamb shift--like phenomena involving $2s_{1/2}$ and $1p_{1/2}$ orbitals, in particular Pauli principle between this last state and the  same state found in the collective quadrupole vibration of the core; c) soft $E1$--mode and related induced pairing.

It could be argued that, in a similar way in which the quadrupole mode renormalize in a very conspicuous way the single--particle orbitals $2s_{1/2}$ and $1p_{1/2}$, it can induce pairing correlations in the $\ket{s^2_{1/2}(0)}$ $\ket{p^2_{1/2}(0)}$ configurations. However, the surface of $^{11}$Li being a misty cloud formed by the halo neutrons, can hardly sustain multipole vibrations any better than the surface of hot nuclei does. Consequently, the only collective mode such a surface can participate in is a dipole mode, in which the negatively charged ($-Z/A\,e\approx-0.27\, e$) neutrons slosh back and forth with respect to the positively charged ($N/A\,e\approx0.73\, e$) protons of the core. Namely, a soft dipole resonance \cite{Kanungo:15} which we also refer to as dipole pygmy resonance (DPR).

In the QRPA calculation of the DPR in $^{11}$Li, the single particle basis associated with $^{10}$Li is worked out making use of a standard parametrized Saxon--Wood potential. The continuum states of this potential are calculated by solving the problem in a spherical box of radius equal to 40 fm, chosen to make the results associated to $^{10}$Li and $^{11}$Li stable. The states at threshold, in particular the parity inverted $s_{1/2}$ and $p_{1/2}$ levels are  the renormalized states, and the amplitudes of the $^{11}$Li ground state wavefunction are used as the $U$, $V$ occupation factors. A separable dipole--dipole interaction ($H_D=-\kappa_1^0 \mathbf D\cdot\mathbf D$) is used with strength $\kappa_1^0\approx -5 V_1/A(R(^{11}\text{Li}))^2$, close to the self--consistent value. Within this scenario one calculates self consistently the full $J^{\pi}=1^-$ spectrum, including the GDR and DPR, fine tuning $\kappa_1^0$ so as to ensure a root at zero energy, which takes care of the elimination of the center of mass motion. The QRPA solutions fulfill the EWSR, the DPR carrying about 10\% of it \cite{Barranco:01}.

If  the virtual $\widetilde{1/2^+}$ state of $^{10}$Li were not at $\approx0.15$ MeV, but considerably higher in energy \cite{Cavallaro:17}, $^{11}$Li would not display the observed properties. But even more, neither the first excited $0^{+*}$ state of $^{12}$Be ($E_x\approx2.25$ MeV), neither the 2.71 MeV, $1^-$ state on top of it will. In fact, these states can be viewed as generated by the new neutron halo pair addition elementary mode of excitation, made out of the symbiotic ($\ket{\tilde 0}_\nu$, DPR) states of $^{11}$Li, acting on $\ket{^{10}\text{Be}(gs)}$. As it emerges from the Figure A5 of \cite{Broglia:16}, such a level scheme is associated with the experimental quantities: $S_{2n}(^{11}\text{Li})=396.6$ keV, $S_{2n}(^{12}\text{Be})=3.672$ MeV, $S_{2n}(^{12}\text{Be})-E_x 0^{+*}=1.422$ MeV, the 2.71 MeV state being likely only part of the dipole state based on $\ket{^{12}\text{Be}(0^{+*})}$. 	

While one has dwelled only on the structure aspects of the one--and two--halo nuclei $^{11}$Be, $^{10}$Li, $^{11}$Li and $^{12}$Be, the associated one-- and two--neutron transfer absolute differential cross sections can also be accurately described within the scenario discussed above, renormalization of single--particle wavefunctions and associated formfactors, playing an important role in the quantitative description of the transfer processes \cite{Barranco:17,Broglia:16,Potel:10,Tanihata:08}.
\section{Conclusions}
The picture of an $s_{1/2}$ state at threshold to create halo nuclei, involves parity inversion, a sizable fraction of the TRK sum rule in low energy $E1$ transitions and absolute value of one-- and two-- particle transfer cross sections of the same order of magnitude, in overall agreement with the experimental findings. At the basis of it one finds the renormalization of single particle motion due to the coupling of the quadrupole vibration of the core, and the pairing induced interaction due to the exchange of the soft $E1$--mode. The soundness of the theory does not stems from a single result, but from the comprehensive picture emerging from the variety of them, in comparison with the data. A single one of these features found incorrect, will set a question mark on the entire approach.

After the above physical arguments and associated estimates collected in the Appendices below had  been written down and worked out, one got hold of the technical detail which likely explains the reason why the data of Cavallaro \textit{et al.} \cite{Cavallaro:17} essentially do not contain any $s_{1/2}$ strength: the angular range in which measurements were carried out \cite{Barranco:18}.

\textit{F. B. acknowledges
funding from the Spanish Ministerio de Econom\'ia under
Grant Agreement No. FIS2017-88410-P. This project has received funding from the European Union’s Horizon 2020 research and innovation program under grant agreement No 654002.}
	\appendix
	\section{}\label{AppA}
	In this Appendix, analytic estimates of structure and reaction properties of the halo nuclei mentioned in the text are provided.
	\subsection{$^{11}_4$Be$_7$, bound one--neutron halo system}
	In the calculation of the structure of $^{11}$Be, four parameters defining the bare single--particle potential $U$ --depth $V$, radius $R$, diffusivity $a$, and spin--orbit strength $V_{ls}$-- were allowed to vary freely so that the  single--particle states  dressed through the coupling to the quadrupole vibration of the core $^{10}$Be, best fitted the data\footnote{It is of notice that the resulting potential to be used with a $k$ mass ($m_k=0.7 m$ (0.9) for $r=0$ ($r=\infty$)), is quite similar to that obtained from SLy4 in the Hartree--Fock approximation.} (ref. \cite{Barranco:17}). The bare single--particle energies $\epsilon_i(i=s_{1/2}, p_{1/2}, d_{5/2})$ are collected in Table \ref{tab1}. Making use of the experimental values of $\hbar\omega_2$ and $\beta_2$ (energy and dynamical quadrupole deformation) of $\ket{^{10}\text{Be}(2^+_1)}$, and the formfactor $R\partial U/\partial r$, the particle--vibration coupling vertices were calculated, and the single--particle states renormalized. The values of the self energies at convergence are shown in Fig. \ref{fig1}, leading to dressed state energies,
	\begin{align}
	\tilde \epsilon_{s_{1/2}}&=(70-570) \text{ keV}=-0.5\text{ MeV},\\
		\tilde \epsilon_{p_{1/2}}&=(-3.04+2.86) \text{ MeV}=-0.180\text{ MeV},\\
			\tilde \epsilon_{d_{1/2}}&=(7.30-1.77-4.08) \text{ MeV}=1.45\text{ MeV}.
	\end{align}
	As seen from Table \ref{tab1}, theory provides an accurate account of the experimental findings. Furthermore, making use of the associated configuration space states
	\begin{align}
	\ket{\widetilde{1/2}^+}=\sqrt{0.80}\,\ket{s_{1/2}}+\sqrt{0.20}\,\ket{(d_{1/2}\otimes 2^+)_{2^+}},
	\end{align}
	and
	\begin{align}
	\ket{\widetilde{1/2}^-}=\sqrt{0.84}\,\ket{p_{1/2}}+\sqrt{0.16}\,\ket{((p_{1/2},p^{-1}_{3/2})_{2^+}\otimes 2^+)_{0^+},p_{1/2}},
	\end{align}
	one obtains, without free parameters
	\begin{align}
	B(E1;1/2^-\to1/2^+)=0.11\,e^2\text{ fm},
	\end{align}
	to be compared with the experimental value
	\begin{align}
	B(E1)=0.102\pm0.002\,e^2\text{ fm},
	\end{align}
	the strongest known electric dipole transition between bound states in nuclei. Comparing this value to the ratio\footnote{Making use of the Thomas--Reiche--Kuhn (TRK) sum rule $TRK=\frac{9}{4\pi}\frac{\hbar^2e^2}{2m}\frac{NZ}{A}\approx 14.8\frac{NZ}{A}e^2\text{ fm}^2$ MeV and of the energy parametrization of the giant dipole resonance (GDR), $\hbar\omega_{GDR}\approx80$ MeV/$(11)^{1/3}\approx36$ MeV, one obtains for $^{11}$Be $TRK/\hbar\omega_{GDR}\approx1e^2\text{ fm}^2$.} $TRK/\hbar\omega_{GDR}\approx1e^2\text{ fm}^2$ one can conclude that the $1/2^-\to1/2^+$ transition carries about 10\% of the TRK (also known as the dipole energy weighted sum rule (EWSR)) and thus about one Weisskopf unit (1$\times B_W(E1)$; see App. \ref{appB}).
	\subsection{$^{11}$Li bound two--neutron halo system}\label{A2}
	In this case we are in presence of a paradigmatic nuclear embodiment of the Cooper pair model. Extending BCS to the single--pair limit, one can estimate the correlation length through the standard relation
	\begin{align}\label{eq7}
	\xi=\frac{\hbar v_F}{\pi|E_{corr}|}\approx20\text{ fm},
	\end{align}
	where use of\footnote{Making use of the Thomas--Fermi model $k_F=(3\pi^2\times 8/(\frac{4\pi}{3}(4.58)^3))^{1/2}\text{ fm}^{-1}\approx0.8\text{ fm}^{-1}$. Thus $(v_F/c)=(\hbar k_F/mc)=0.2\times\text{ fm }\times k_F\approx0.16$.} $(v_F/c)\approx0.16$ and $E_{corr}\approx-0.5$ MeV was made. Dividing the density distribution of nucleons in $^{11}$Li into a compact, normal closed shell $N=6$ core $^{9}$Li of radius $R_0=1.2(9)^{1/3}$ fm$\approx2.5$ fm, and a Cooper pair of correlation length (\ref{eq7}), one can work out a simple estimate of the effective radius of $^{11}$Li as,
	\begin{align}
	R_{eff}=\left(\frac{9}{11}\times(2.5)^2+\frac{2}{11}\left(\frac{\xi}{2}\right)^2\right)^{1/2}\approx 4.8\text{ fm},
	\end{align}
	leading to $\langle r^2\rangle^{1/2}=\sqrt{\frac{3}{5}}R_{eff}\approx3.7$ fm, to be compared with $\langle r^2\rangle_{exp}^{1/2}=3.55\pm0.1$ fm. 
	
	Let us now work out a simple estimate of $E_{corr}$ used in (\ref{eq7}). There is experimental evidence \cite{Thoennessen:99,Zinser:95,Chartier:01,Simon:07} of the presence in $^{10}$Li, of a $1/2^+$ virtual state and of a low--lying $1/2^-$ resonant state\footnote{See also \cite{Cavallaro:17} and \cite{Barranco:18}.}. In keeping with the analytic results of Sect. \ref{A3} (see below), we assume these states to be $\ket{\widetilde{1/2}^+}=\ket{\widetilde{s_{1/2}; 0.15\text{ MeV}}}$, and $\ket{\widetilde{1/2}^-}=\ket{\widetilde{p_{1/2}; 0.60\text{ MeV}}}$. Again the scenario of a low--lying collective, soft $E1$--mode.
	
	Making the ansatz of transferability from $^{11}$Be, one can ascribe to this soft mode, a 10\% of the TRK sum rule to solve the RPA dispersion relation
	\begin{align}\label{eq10}
	W(E)=\sum_{ki}\frac{2(\epsilon_k-\epsilon_i)|\braket{i|F|k}|^2}{(\epsilon_k-\epsilon_i)^2-E^2}=\frac{1}{\kappa_1^0}.
	\end{align}
	In the schematic calculations carried out here, the full quasiparticle subspace discussed in Sect. \ref{S2.2} in connection with the QRPA calculation of the DPR is reduced to the $\tilde s_{1/2}\to\tilde p_{1/2}$ transition, and only the neutron halo degrees of freedom are considered. Degrees of freedom which constitute (2/11) of the total nucleonic space and feels an effective confining radius $\xi/2$. In other words, the factor (1/$R_{eff}^2$) entering in $\kappa_1^0$ is to be replaced in the present estimates by $\frac{1}{\left(\frac{\xi}{2}\right)^2}\times\left(\frac{2}{11}\right)$. It  leads to a dipole screened coupling constant of value $\kappa_1=\frac{R_{eff}^2}{\left(\frac{\xi}{2}\right)^2}\kappa_1^0\approx 0.04\kappa_1^0\approx-0.021$ MeV fm$^2$,
	where 
	\begin{align}
	\kappa_1^0=-\frac{5V_1}{AR_{eff}^2}\quad,\quad V_1=25\text{ MeV}.
	\end{align}
	Replacing $k$ and $i$ by the renormalized states $\ket{\widetilde{1/2}^-}$ and $\ket{\widetilde{1/2}^+}$, and $E$ by the energy of the dipole pygmy resonance (DPR) to be determined, one can write
	\begin{align}
	(\tilde\epsilon_k-\tilde\epsilon_i)^2-(\hbar\omega_{DPR})^2=\kappa_1\times 2\times(0.1\times TRK),
	\end{align}
	where\footnote{Associated with the operator $F(r_k)=e\left[\frac{N-Z}{2A}-t_z(k)\right]r_k$ $(t_z(k)=\pm1/2)$, and thus no spherical harmonic.\label{f7}}
	\begin{align}\label{eq13}
	TRK=\frac{3\hbar^2}{2m}\frac{NZ}{A}=131\text{ MeV fm}^2\quad (^{11}_3\text{Li}_8).
	\end{align}
	Thus
	\begin{align}
\nonumber	\hbar\omega_{DPR}&=((0.6-0.15)^2\text{ MeV}^2-(-0.021\text{ MeV fm}^{-2})\\
	&\times2\times0.1\times131\text{ MeV fm}^2)^{1/2}\approx (0.45^2+0.74^2)^{1/2}\approx 1\text{ MeV},
	\end{align}
	as compared to the centroid value of the resonance observed in $d(^{11}\text{Li},d')$ experiment leading to $1.03\pm0.03$ MeV \cite{Kanungo:15}.
	
	Let us now calculate the particle vibration coupling (PVC) strength $\Lambda$ of this mode to the nucleons. Note the use in the following estimates of a dimensionless dipole single particle field $F'=F/R_{eff}(^{11}$Li). This is in keeping with the fact that one aims at obtaining a quantity with energy dimensions ($[\Lambda]=$ MeV), and $\kappa_1^0$ has been introduced in Eq. (\ref{eq10}) as the self consistent value of the dipole--dipole separable interaction, normalized in terms of $R^2_{eff}$ ($^{11}$Li). An alternative way to obtain a similar result, is to work out the value of $\Lambda$ without the ($1/R_{eff}^2$), and multiply the result  by $\overline{|\braket{i|F|k}|^2}$. It is expected that both results agree within 10-20\% effects. One then obtains,
		\begin{align}
\nonumber\Lambda^2&=\left(\left(\frac{\partial W'(E)}{\partial E}\right)_{\hbar\omega_{DPR}}\right)^{-1}=\left\{2\hbar\omega_{DPR}\frac{2\times0.1\times TRK/R^2_{eff}}{\left[(\tilde\epsilon_{p_{1/2}}-\tilde\epsilon_{s_{1/2}})^2-(\hbar\omega_{DPR})^2)\right]^2}\right\}^{-1}\\
&=\left(\frac{2.3}{0.64\text{ MeV}^2}\right)^{-1}\approx0.28\text{ MeV}^2\quad (\Lambda=0.53\text{ MeV}).
		\end{align}
 The induced pairing interaction associated with the exchange of the DPR between the two halo neutrons leads to the matrix element
\begin{align}
M_{ind}\approx-\frac{2\Lambda^2}{\hbar\omega_{DPR}}\approx-0.6\text{ MeV},
\end{align} 	
	the factor of 2 being associated with the two possible time orderings (Fig. \ref{fig2}). Let us now calculate the bare pairing interaction, taking into account the screening\footnote{The estimate $G\approx28\text{ MeV}/A$ of the pairing strength is made with the help of a $\delta$--force \cite{Brink:05}. The corresponding matrix element in the configuration $j^2(0)$ can be expressed as $\braket{j^2(0)|V_\delta|j^2(0)}=-\frac{2j+1}{2}G$, with $G=V_0/R_0^3\approx28/A\text{ MeV}(R_0=1.2 A^{1/3}$ fm). Consequently in the case of $^{11}$Li the strength $G$ will be screened by the factor ($2/(2j+1))\mathcal O$ where $\mathcal O=\left(\frac{R_0(^{11}\text{Li})}{R_{eff}}\right)^3$ is the overlap between the core and the halo wavefunctions, and $j=k_FR_0=1.36\text{ fm}^{-1}\times 2.7\text{ fm}\approx3.7$ and $2j+1\approx8$, while $2j+1=2$ for both $s_{1/2}$ and $p_{1/2}$.} due to the large radius of $^{11}$Li. That is
	\begin{align}
	(G)_{scr}=\frac{2}{8}\left(\frac{2.7}{4.8}\right)^3G\approx0.045\times\frac{28}{A}\text{ MeV}\approx\frac{1.3\text{ MeV}}{A}\approx0.1\text{ MeV}.
	\end{align}
	Consequently, the neutron halo Cooper pair binds the core $^9$Li, with the correlation energy
	\begin{align}
	E_{corr}=2\tilde\epsilon_{s_{1/2}}-(G)_{scr}+M_{ind}=(0.3-0.1-0.6\text{ MeV})\approx-0.4\text{ MeV},
	\end{align}
to be compared with the experimental value $(E_{corr})_{exp}=-380$ keV.

Furthermore, the absolute cross sections associated with the $^1$H($^{11}$Li, $^9$Li($\mathbf f$)$^3$H reaction, requires two groups of components sharing about evenly the normalized value of the sum of the squared amplitudes. A many--body particle--hole--like one $\alpha\ket{(p_{1/2},s_{1/2})_{1^-}\otimes 1^-;0^+}+\beta\ket{(s_{1/2},d_{5/2})_{2^+}\otimes 2^+;0^+}$\\ \mbox{$(\mathbf f=1/2^-, 2.69\text{ MeV}; \alpha^2+\beta^2\approx0.5,\alpha\gg\beta)$}, and another pairing--like\\
 $\gamma\ket{s^2_{1/2}(0)}+\delta\ket{p^2_{1/2}(0)}$ ($\mathbf f=gs,\gamma^2+\delta^2\approx0.5,\gamma\approx\delta$). Thus, concerning this second one, one can write $\ket{0}=\sqrt{0.25}\ket{s^2_{1/2}}+\sqrt{0.25}\ket{p^2_{1/2}}$. If the first component was to be set equal to zero and thus, because of normalization $\sqrt{0.5}\ket{p^2_{1/2}}$, the absolute two--particle ground state cross section will be predicted a factor $\approx7$ smaller than that associated with $\ket{0}$, which reproduces the observed absolute differential cross section, within experimental errors \cite{Potel:10,Tanihata:08}. This in keeping with the fact that $\sigma(s^2_{1/2}(0))\approx15$ mb, while $\sigma(p^2_{1/2}(0))\approx2$ mb. Thus $\sigma(\sqrt{0.25}\sqrt{14}+\sqrt{0.25}\sqrt{2})^2$ mb $\approx6.7$ mb, while $(\sqrt{0.5}\times\sqrt{2})^2$ mb $\approx1$ mb. 
\subsection{$^{10}$Li, unbound one--neutron halo system: structure and reactions in the continuum}\label{A3}
In this case we use as input the value of $\hbar\omega_{2^+}\approx3.3$ MeV and $\beta_2=0.8$ characterizing the low--lying quadrupole vibration of the core $^9$Li, as well as $R_{eff}(^{11}$Li)=4.8 fm worked out in the previous section.

As bare potential we use the standard WS potential $U(r)=Uf(r), f(r)=\left(1+\exp\left(\frac{r-R_0}{a}\right)\right)^{-1}$, where $U=U_0+0.4E$, $U_0=V_0+30(N-Z)/A$ MeV and  $V_0=-51$ MeV. The energy dependent term ($E=\hbar k^2/2m-\epsilon_F$) is taken care of by the $k$--mass $m_k=(1+0.4\times\mathcal O)^{-1}m\approx 0.93 m$, where the overlap between halo and core single--particle wavefunctions is $\mathcal O=\left(2.7/4.8\right)^3\approx 0.2$, as defined in Sect. \ref{A2}. Expressed differently, because of the large radius of the halo, the Pauli principle plays little role in the mean field, and $m_k\approx m$, $m$ being the bare nucleon mass. Making use of the above potential and of the associated symmetry and spin--orbit terms, the bare single--particle energies $\epsilon_{p_{3/2}}, \epsilon_{p_{1/2}},\epsilon_{s_{1/2}}$ and $\epsilon_{d_{5/2}}$   were calculated. They are displayed in Table \ref{tab2}.

\begin{figure}[h]
		\centerline{\includegraphics*[width=15cm,angle=0]{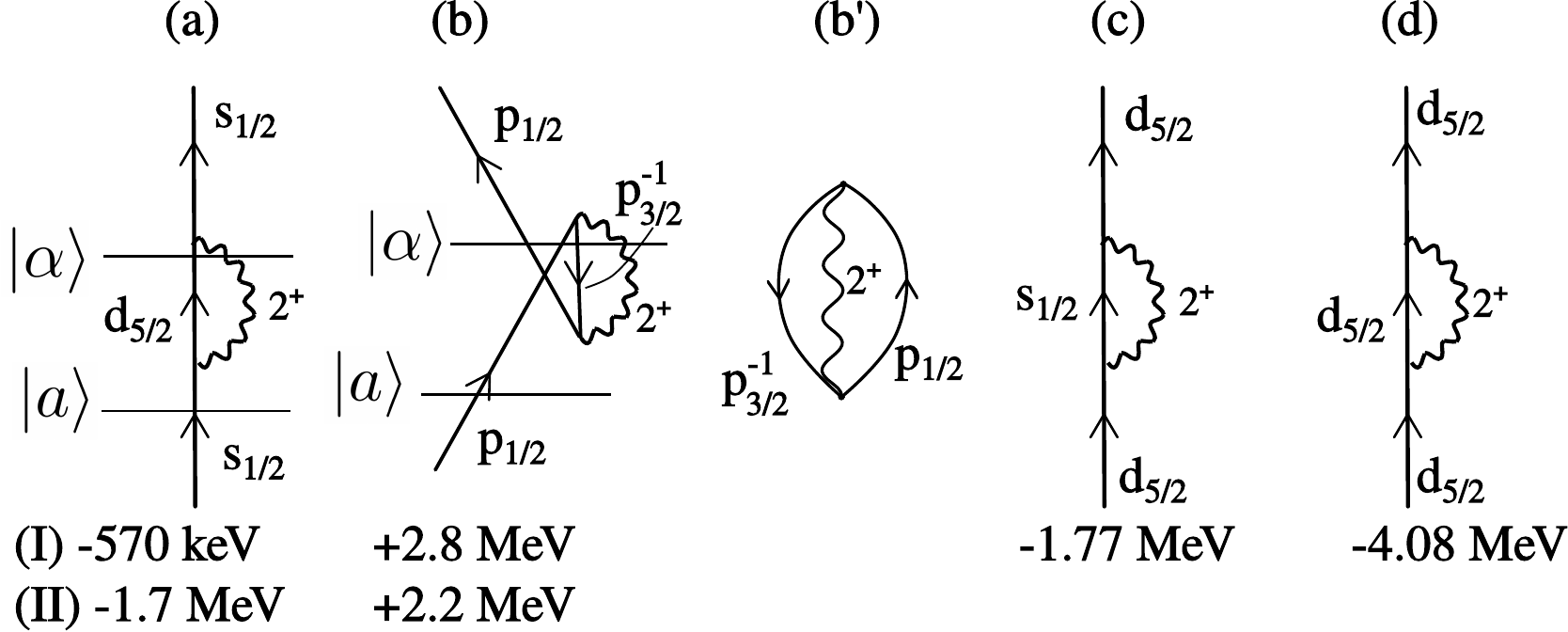}}
		\caption{Self--energy associated with the lowest $s_{1/2}$ (\textbf{a}) $p_{1/2}$ (\textbf{b}) states and $d_{5/2}$ (\textbf{c}) and (\textbf{d}) resonance of $^{11}$Be (line (\textbf{I})) and of the $s_{1/2}$ virtual and $p_{1/2}$ resonant states of $^{11}$Li (line (\textbf{II}).) The diagram (\textbf{b}') describes the ZPF associated with the component ($p_{1/2},p_{3/2}^{-1})_{2^+}$ of the quadrupole mode of the core $^9$Li.. The label $\ket{a}$ stands for the state to be renormalized due to the coupling to the intermediate (virtual) state $\ket{\alpha}$. It is to be noted that the results displayed in (I) are at convergence, i.e. obtained by summing to all orders the corresponding process \cite{Barranco:17}, while those shown in (II) are second order in the PVC vertex results (see \ref{eq21} and \ref{eq30}).}\label{fig1}
	\end{figure}

\begin{figure}[h]
		\centerline{\includegraphics*[width=8cm,angle=0]{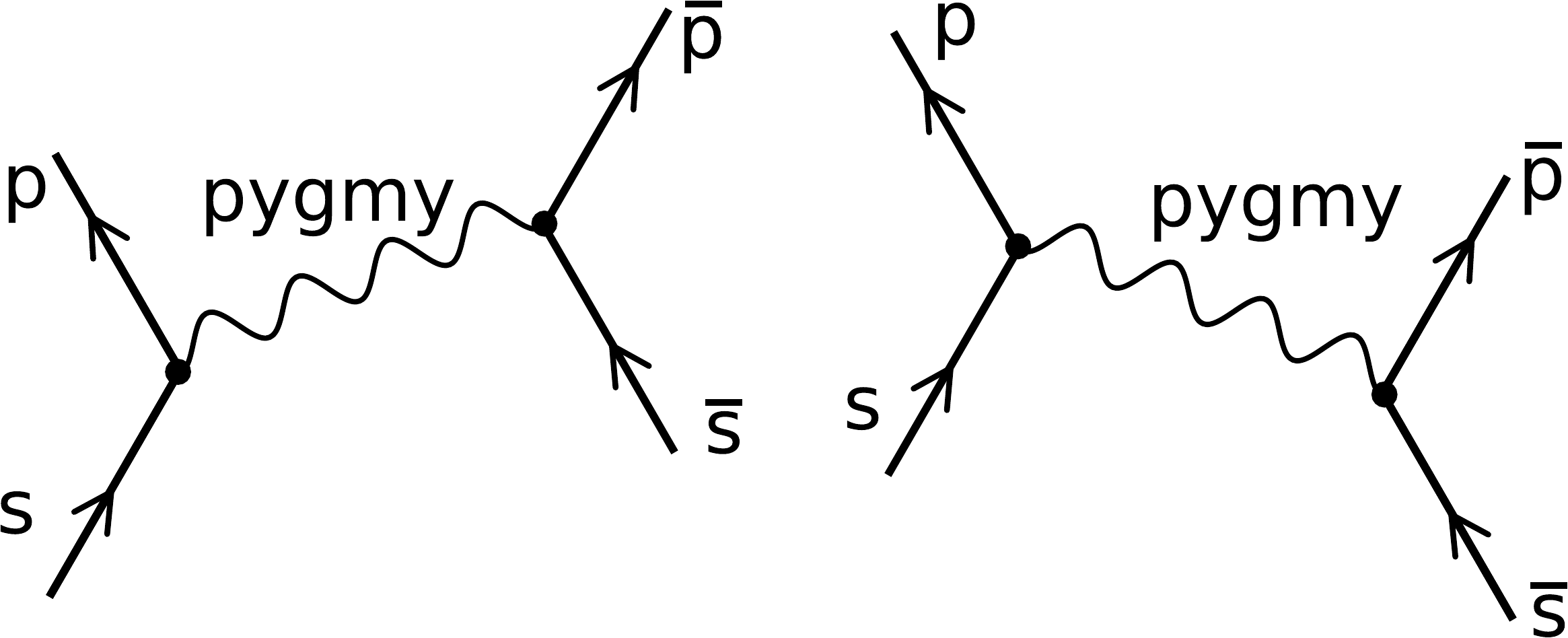}}
		\caption{Induced pairing interaction between the halo neutrons resulting from the exchange of the DPR between the configurations $s^2(0)$ and $p^2(0)$. It is of notice that $\ket{\bar \nu}$ stands for the state time reversed to $\ket{\nu}$ ($\nu=s,p$).}\label{fig2}
	\end{figure}

With the help of  (-51+30$\frac{6-3}{9}$) MeV=-41 MeV, and of $\braket{R_0\partial U/\partial r}\approx 1.44\times U_0\approx-60$ MeV (\cite{Brink:05}, App. D), one can calculate the PVC vertex associated with the quadrupole vibration of the core,
\begin{align}
\nonumber v&=\braket{H_c}=\frac{\beta_2}{\sqrt{5}}\braket{R_0\frac{\partial U}{\partial r}}\mathcal O \braket{j||Y_2||1/2},\\
&\braket{j||Y_2||1/2}=\sqrt{\frac{2j+1}{4\pi}}=\left\{\begin{array}{cc}
0.7&j=5/2 (d_{5/2})  \\ 
0.6& j=3/2 (p_{3/2}),
\end{array} \right.\\
&v\approx\frac{0.8}{\sqrt{5}}(-60\text{ MeV})\times0.2\times0.7\approx-3\text{ MeV}.\label{eq24}
\end{align}
In keeping with the fact that the $p_{1/2}$ is a bound state while $s_{1/2}$ is not, the corresponding wavefunction is more concentrated and, consequently, the corresponding matrix elements of $H_c$ larger. We take the empirical ratio 26/20$\approx1.3$ between $G_p$ and $G_n$ as indicative (\cite{Brink:05}, p. 63). In what follows we shall thus use $v_{s_{1/2}}\approx -3.0$ MeV and $v_{p_{1/2}}\approx -3.9$ MeV.
	Let us now calculate the renormalization of the $s_{1/2}$ and $p_{1/2}$ states. The self--energy diagram (a) of Fig. \ref{fig1} gives, to second order of perturbation in $v$,
\begin{align}\label{eq21}
\Sigma_{s_{1/2}}=\frac{v^2}{\epsilon_{s_{1/2}}-(\epsilon_{d_{5/2}}+\hbar\omega_2)}=\frac{9\text{ MeV}^2}{(1.5-6.8)\text{ MeV}}=-1.7\text{ MeV.}
\end{align}

	\begin{table}[h]
		\begin{center}
			\begin{tabular}{|c  | c | c | c |}
				\hline
				$i$&$\epsilon_i$ (MeV)&$\tilde\epsilon_i$ (MeV)& $(\epsilon_i)_{exp}$ (MeV)$^{\text{b)}}$\\
				\hline
				$s_{1/2}$&0.07&-0.5& -0.5\\
				$p_{1/2}$&-3.04&-0.18& -0.18\\
				$d_{5/2}$&7.30&1.45& 1.28$^{\text{a})}$\\
				\hline
			\end{tabular}
		\end{center}
		$^{\text{a)}}$ Centroid of resonance\\
		$^{\text{b)}}$ \cite{Winfield:01} (see also \cite{Barranco:17} and references therein).
		\caption{$^{11}$Be: bare $(\epsilon_i)$, dressed ($\tilde{\epsilon}_i$), and experimental $\epsilon_{exp}$ single--particle energies of the lowest bound and resonant states.}\label{tab1}
	\end{table}
	
	\begin{table}[h]
		\begin{center}
			\begin{tabular}{|c  | c | c | c |}
				\hline
				$i$&$\epsilon_i$ (MeV)&$\tilde\epsilon_i$ (MeV) &$(\epsilon_i)_{exp}$ (MeV)$^{\text{a),b)}}$\\
				\hline
				$d_{5/2}$&3.5& &  \\
				$s_{1/2}$&1.2&0.15$^{\text{c)}}$& 0.1-0.25$^{\text{a)}}$\\
				$p_{1/2}$&-1.2&0.60$^{\text{d)}}$& 0.4-0.6$^{\text{a),b)}}$\\
				$p_{3/2}$&-4.7&&\\
				\hline
			\end{tabular}
		\end{center}
		$^{\text{a)}}$ \cite{Zinser:95}\\			
		$^{\text{b)}}$ \cite{Cavallaro:17}\\
		$^{\text{c)}}$ virtual\\
		$^{\text{d)}}$ resonant\\	
		\caption{Same as Table \ref{tab1}, but for $^{10}$Li.}\label{tab2}
	\end{table}

	The renormalized energy $\tilde \epsilon_{s_{1/2}}$ at convergence is obtained by solving the secular equation
	\begin{align}\label{eq26} 
\left|\begin{array}{cc}
(E_a-E_i)& v\\ 
v & (E_\alpha-E_i) 
\end{array}\right| = \left|\begin{array}{cc}
(1.5-E_i) & -3.0\\ 
-3.0 & (6.8-E_i) 
\end{array}\right|=0.	
	\end{align}
	That is
	\begin{align}\label{eq27}
	E_i^2-8.3E_i+1.2=0,
	\end{align}	
	where all the numbers in (\ref{eq26}) and (\ref{eq27}) are in MeV. The lowest root of (\ref{eq27}) is $E_1=\tilde \epsilon_{s_{1/2}}=0.15$ MeV. It is of notice that in the present case, as well as in connection with the calculation of $\tilde \epsilon_{p_{1/2}}$ below, perturbation theory i.e. $\epsilon_{s_{1/2}}+\Sigma_{s_{1/2}}=(1.5-1.7)$ MeV =-0.2 MeV cannot be used, and the process displayed in Fig. \ref{fig1} (a) has to be summed to all orders of perturbation. 
	
	The above provides a textbook example of the specificity with which one can single out, within the framework of NFT, the physical processes at the basis of a phenomenon under study, e.g. parity inversion in $^{10}$Li, and the economy with which one can ``exactly'' treat them. But also only them, not being forced to waste resources, but most importantly, physical insight in keeping track at the same time of myriads of little relevant but somewhat connected processes.
	
	Let us now work out the amplitudes of the $\ket{\tilde \epsilon_{1/2}}$ state. That is
	\begin{align}\label{eq28}
\nonumber 	c^2_{s_{1/2}}(1)&=\left(1+\frac{v^2}{(E_\alpha-E_1)^2}\right)^{-1}=\left(1+\frac{9\text{ MeV}^2}{(6.8-0.15)^2\text{ MeV}^2}\right)^{-1}\\
	&=(1+0.2)^{-1}=0.83.
	\end{align}
	Making use of normalization ($c^2_a(1)+c^2_\alpha=1$) one can write
	\begin{align}
	\ket{\widetilde{1/2^+;0.15}}=0.91\ket{s_{1/2}}+0.41\ket{(d_{5/2}\otimes2^+);1/2^+}
	\end{align}
	From  (\ref{eq28}) one obtains that the mass enhancement factor is $\lambda=0.2$ and thus the effective $\omega$--mass $m_\omega=1.2m$, while the discontinuity at the Fermi energy (single--particle content) is $Z_\omega=\left(m_\omega/m\right)^{-1}=0.83$.
	
	Let us now discuss the renormalization of the $p_{1/2}$ state. Following the same steps as before we find
\begin{align}\label{eq30}
\nonumber \Sigma_{p_{1/2}}=&\frac{(-1)^1(-3.9\text{ MeV})^2}{\epsilon_{p_{1/2}}-\left[(2\epsilon_{p_{1/2}}-\epsilon_{p_{3/2}})+\hbar\omega_2\right]}=\frac{15.21\text{ MeV}^2}{\left[\epsilon_{p_{1/2}}-(\epsilon_{p_{3/2}}-\hbar\omega_2)\right]}\\
&=\frac{15.21\text{ MeV}}{-1.2-(-4.7-3.3)}=\frac{15.21}{-1.2-(-8)}=\frac{15.21}{6.8}=2.2\text{ MeV},
\end{align}	
	where the phase $(-1)^n$ comes from the number of unavoidable crossing ($n=1$) in diagram (b) of Fig. \ref{fig1}. Thus, in second order perturbation theory, $\epsilon_{p_{1/2}}+\Sigma_{p_{1/2}}=-1.2$ MeV +2.2 MeV=1.0 MeV. 
	
	Let us now calculate the same process (a) Fig. \ref{fig1} to convergence. For this we need $E_a=\epsilon_{p_{1/2}}=-1.2$ MeV, $E_\alpha=\epsilon_{p_{3/2}}-\hbar\omega_{2^+}=(-4.7-3.3)$ MeV= -8 MeV. The associated secular equation being
	\begin{align}\label{eq32} 
	\left|\begin{array}{cc}
	(-1.2-E_i)& -3.9\\ 
	-3.9 & (-8-E_i) 
	\end{array}\right| = E_i^2+9.2E_i-5.61=0,	
	\end{align}	
	again all numbers in MeV. We then obtain $E_1=\tilde \epsilon_{p_{1/2}}=0.6$ MeV, a result which again testifies to the inapplicability of perturbation theory. The square amplitude of the corresponding renormalized state is
	\begin{align}\label{eq33} 
	c^2_{p_{1/2}}=\left(1+\frac{(-3.9\text{ MeV})^2}{(0.6-(-8))^2}\right)^{-1}=(1+0.21)^{-1}=0.83.
	\end{align}
	One then can write
	\begin{align}
	\ket{\widetilde{1/2^-;0.6\text{ MeV}}}=0.91\ket{p_{1/2}}+0.41\ket{((p_{1/2},p^{-1}_{3/2})_{2^+}\otimes2^+)_{0^+}p_{1/2};1/2^-},
	\end{align}	
	the associated mass enhancement factor, $\omega$--mass and $Z_\omega$--factor being $\lambda=0.21$, $m_\omega=1.21m$ and $Z_\omega=0.83$ respectively.	
	The energy of the parity inverted  states are compared in Table \ref{tab1} with the experimental findings. 
	
	In connection with the energy associated with the intermediate state $E_\alpha=\epsilon_{p_{3/2}}-\hbar\omega_{2^+}=(-4.7-3.3)$ MeV=-8 MeV in  Eqs (\ref{eq32}) and (\ref{eq33}), we refer to the energy denominator of Eq. (\ref{eq30}) for a mathematical explanation. The physics can be found in the Lamb shift--like effect described by diagram (b) of Fig. \ref{fig1}, a phenomenon closely connected with Pauli principle and the ZPF process shown in diagram (b'). To allow the dressed $\ket{\widetilde p_{1/2}}$ state to acquire asymptotic waves, i.e. to be on shell, one has to annihilate simultaneously the quadrupole phonon and the $p^{-1}_{3/2}$ hole implying  an overall energy change of -3.3 MeV-4.7 MeV=-8 MeV. 
	\section{$E1$--Weisskopf unit}\label{appB}
	The $E1$--unit, so called Weisskopf unit is defined\footnote{Both in the definition of $B_W$ as well as of $S(E1)$ below (Eq. (\ref{eq30})) and at variance to TRK (Eq. (\ref{eq13}), see also footnote \ref{f7}), the corresponding dipole operator contains the spherical harmonics of multipolarity $\lambda=1$.} as $B_W(E1)=((1.2)^2/4\pi)(3/4)^2A^{2/3}e^2$ fm$^2=\frac{0.81}{4\pi}A^{2/3}e^2$ fm$^2$ (\cite{Bohr:69} p. 389, Eq(3C-38)), which together with the comment at the end of p. 387 and starting of p. 388,
	\begin{align}
	e\to(e)_{E1}=\left\{\begin{array}{cc}
	\frac{N}{A}e=\frac{8}{11}e=0.73e&  \text{protons}\\  
	-\frac{Z}{A}e=-\frac{3}{11}e=-0.27e&  \text{neutrons} 
	\end{array} \right.
	\end{align}
	implies, for $^{11}_3$Li$_8$
	\begin{align}
	B_W(E1)=0.32(e)^2_{E1}\text{ fm}^2=\left\{\begin{array}{cc}
	0.17e^2\text{ fm}^2&  \text{(p)}\\  
	0.023e^2\text{ fm}^2&  \text{(n)}, 
	\end{array} \right.
	\end{align}	
	and thus an average value
	\begin{align}
	\overline{B_W(E1)}=\frac{0.17+0.023}{2}e^2\text{ fm}^2\approx0.1 e^2\text{ fm}^2
	\end{align}
	Making now use of \cite{Bohr:75} p. 403 Eq. (6-176) for the case of $^{11}$Li,
	\begin{align}\label{eq34}
	S(E1)=14.8\frac{NZ}{A}e^2\text{ fm}^2\text{ MeV}=32.3 e^2\text{ MeV fm}^2
	\end{align}
	together with $\hbar\omega_{GDR}=80$ MeV$/A^{1/3}$ MeV$\approx36$ MeV gives
\begin{align}
\frac{S(E1)}{\hbar\omega_{GDR}}\approx 0.9 e^2\text{ fm}^2.
\end{align}
	Assuming the DPR of $^{11}$Li to carry $\approx$10\% of $S(E1)$ one then obtains
	\begin{align}
	10\%\left(\frac{S(E1)}{\hbar\omega_{GDR}}\right)\approx0.09 e^2\text{ fm}^2,
		\end{align}
	and thus a Weisskopf unit.
	

 \end{document}